\documentclass[11pt,a4paper]{article}
\usepackage[utf8]{inputenc}
\usepackage[T1]{fontenc}
\usepackage[margin=25mm]{geometry}
\usepackage{graphicx}
\usepackage{amsmath,amssymb}
\usepackage{booktabs,tabularx,array,multirow}
\usepackage[super]{cite}
\usepackage{url}
\usepackage{enumitem}
\usepackage{caption}
\usepackage{booktabs}
\usepackage{tabularx}
\usepackage{array}

\renewcommand{\arraystretch}{1.2}
\newcolumntype{L}[1]{>{\raggedright\arraybackslash}p{#1}}
\newcolumntype{Y}{>{\raggedright\arraybackslash}X}
\newif\ifembedfigures
\embedfigurestrue         

\newcommand{\figureasset}[2]{%
  \ifembedfigures
    \IfFileExists{#1}{\includegraphics[width=0.95\linewidth,height=80mm,keepaspectratio]{#1}}{%
      \fbox{\parbox[c][45mm][c]{0.92\linewidth}{\centering
      \textbf{Figure placeholder}\par\smallskip\small\itshape #2\par\smallskip
      \footnotesize\upshape File: \texttt{\detokenize{#1}}}}}%
  \fi}

\title{MM-CleanJerk: A Synchronized Multimodal Wearable-IMU and Dual-View Video Dataset of Olympic Clean and Jerk Lifting with Phase Annotations}

\author{%
  \normalsize Thevindu Samarasekera$^{1,*}$, Janith Yapa$^{1}$,
  Praveen Rathnayaka$^{2}$, Roshan Godaliyadda$^{1,2}$,\\[0.25em]
  \normalsize Parakrama Ekanayake$^{1,2}$, Chanaka Senanayake$^{3}$,
  Asela Ratnayake$^{4}$%
}

\date{}

\begin{document}
\maketitle
\vspace{-1.5em}
\begin{center}\small
$^1$[Multidisciplinary AI Research Centre, University of Peradeniya, Peradeniya, Sri Lanka]\par
$^2$[Department of Electrical and Electronic Engineering, Faculty of Engineering, University of Peradeniya, Peradeniya, Sri Lanka]\par
$^3$[Department of Manufacturing and Industrial Engineering, Faculty of Engineering, University of Peradeniya, Peradeniya, Sri Lanka] \par
$^4$[National Hospital of Kandy, Kandy, Sri Lanka] \par
$^*$Correspondence: [Thevindu Samarasekera at thevindu.s@eng.pdn.ac.lk]\par
\end{center}

\begin{abstract}
Two clean and jerk lifts can take the same time even when an athlete spends longer recovering from the clean in one lift and stabilising the bar overhead in the other. MM-CleanJerk captures this variation across 246 lifts performed in 54 sets by 19 athletes during coached training in Sri Lanka. Participants ranged from novices to national competitors. We recorded acceleration and angular velocity at 100 Hz using 26 inertial measurement units on the limbs and spine. Front and side videos recorded at 30 frames per second were synchronised with the sensor signals and are available for 28 sets. The median absolute residual lag after synchronisation was 31.2 ms. Three annotators independently marked 12 events defining 11 phases in each lift. Across all events, the median standard deviation of their timestamps was 69.4 ms. The release includes individual annotations, consensus markers and barbell loads, allowing researchers to compare phase timing across repeated lifts and evaluate methods for detecting transitions from wearable sensors.
\end{abstract}


\section{Background \& Summary}
\label{sec:background and summary}

Two clean and jerk lifts of equal duration can differ in how the athlete receives and stabilises the barbell. An athlete may spend longer with the bar at the shoulders before initiating the jerk in one lift and longer regaining balance overhead in the next. A repetition count or a single duration conceals this distinction. To compare the execution of repeated lifts, researchers need to identify the same events across repetitions and relate those events to the motion of individual body segments. We present a dataset collected during coached weightlifting sessions in Sri Lanka that links whole-body inertial recordings to video-derived phase boundaries for this purpose.

Body-worn inertial measurement units (IMUs) record acceleration and angular velocity throughout a movement and can be used in the athlete's usual training environment \cite{garcia2023,liang2023}. Whereas optical systems require a calibrated recording volume and a clear view of the athlete, a body-worn sensor continues to record when a limb passes behind the bar or trunk \cite{garcia2023}. This makes inertial measurements attractive for technique assessment in sport, where the practical challenge is to relate a change in a sensor signal to a specific action by the athlete \cite{souaifi2025}. The information available depends on where the sensors are placed. An upper-arm sensor, for example, has been used to identify and count overhead throws and serves \cite{rawashdeh2016}. In lifting exercises, Michaud et al. used two sensors to measure relative rotation across the lumbopelvic region during deadlifts and kettlebell swings \cite{michaud2021}. Conforti et al. used eight IMUs to examine lower-limb and trunk kinematics during manual handling and found that hip and trunk motion changed with load \cite{conforti2020}. Whole-body inertial recordings have also been used as inputs to inverse dynamics models that estimate joint moments during manual handling \cite{diraneyya2021}. For reuse, it therefore matters whether a dataset supplies the original measurements, the locations at which they were acquired and the external load for each recording. A pair of sensors can address a local question such as hip-hinge execution. Examining how the legs, trunk and arms coordinate across the clean and the jerk requires measurements at several anatomical sites on a common time base.

Video analyses of competitive weightlifting have examined how barbell speed and the duration of individual movement phases vary across the clean and jerk \cite{ulareanu2014}. Relating those observations to body-worn measurements poses a timing problem. During the clean, the athlete raises the bar, moves under it, receives it at the shoulders and recovers to standing. The jerk begins from this front rack position and includes a dip, an upward drive and an overhead reception. A downward movement of the body can therefore occur both when receiving the clean and when initiating the jerk. The preceding bar position and the motion of adjacent segments help distinguish these events. Frontal and sagittal views provide complementary evidence of bar position, foot placement and recovery, while IMUs follow segment motion through the transitions. A further challenge is deciding exactly when a moving athlete becomes stationary. In the present recordings, the median standard deviation across annotators was 50.9~ms for the first-pull marker and 117.1~ms for the post-lift rest marker. Retaining the individual annotations makes this difference visible to users studying event timing.

Existing datasets address complementary aspects of the measurement problem (Table~\ref{tab:background-datasets}). CeTI-Age-Kinematics documents repeated daily tasks with whole-body and hand measurements \cite{pogrzeba2025}. Scherpereel et al. recorded lower-limb biomechanics across activities ranging from walking to jumping and squatting \cite{scherpereel2023}. GAITEX pairs inertial and optical recordings with temporal segmentations and movement-quality ratings for gait and rehabilitation exercises \cite{spilz2026}. In resistance exercise, MM-Fit links wearable signals to video across workout sessions for exercise recognition and repetition counting \cite{stromback2020}, while Velloso et al. investigated how wearable measurements distinguish differences in exercise execution \cite{velloso2013}. In sport, MultiSenseBadminton links individual strokes to video, sensor measurements and annotations across skill levels \cite{seong2024}. These resources provide examples of exercise recognition, technique assessment and temporal annotation. A question specific to Olympic weightlifting is how much time elapses between successive events in a loaded lift and which segments move during that interval. The present collection links each event to the corresponding video frame and inertial sample. It preserves the surrounding set so that researchers can study how those intervals vary across consecutive lifts by the same participant.

\begin{table}[htbp]
\centering\small
\caption{Selected movement datasets and the measurements or annotations relevant to the comparison.}
\label{tab:background-datasets}
\begin{tabularx}{\linewidth}{@{}>{\raggedright\arraybackslash}p{0.27\linewidth}>{\raggedright\arraybackslash}X>{\raggedright\arraybackslash}X@{}}
\toprule
Dataset or study & Recorded tasks & Measurements or annotations \\
\midrule
CeTI-Age-Kinematics \cite{pogrzeba2025} & 30 daily tasks, 32 younger and older adults & Inertial suits and gloves, repeated tasks and anthropometry \\
Scherpereel et al. \cite{scherpereel2023} & 20 non-cyclic and 11 cyclic activities, 12 participants & Lower-limb kinematics, force plates, IMUs and electromyography \\
GAITEX \cite{spilz2026} & Gait and rehabilitation exercises, 19 participants & Nine IMUs, optical markers, temporal segments and quality ratings \\
Yasar et al. \cite{yasar2024} & Repeated shoulder rotations to fatigue, 34 participants & Inertial and physiological signals, load relative to voluntary strength and perceived exertion \\
MM-Fit \cite{stromback2020} & Full-body workout sessions & Consumer wearables, colour and depth video, and pose estimates \\
MultiSenseBadminton \cite{seong2024} & Forehand clear and backhand drive strokes, 25 players & Body tracking, video, stroke labels and skill levels \\
This dataset & 246 clean and jerk lifts in 54 sets, 19 participants & 26 IMU sites, 12 phase markers per lift, individual annotations and two-view video for selected sets \\
\bottomrule
\end{tabularx}
\end{table}

The acquisition platform builds on our earlier work on synchronised wearable instrumentation for knee and elbow motion \cite{adhikari2026,samarasekera2026}. That work examined whether the platform could resolve changes in joint motion during controlled tests and clean and jerk lifting. Here the contribution is the recording collection itself, including the measurements, anatomical sensor map, set-level loading information and annotations needed to investigate questions beyond the original instrumentation study.

The dataset comprises 246 lifts performed by 19 participants across 54 retained sets at the University of Peradeniya Gymnasium. Participants were recruited through the Sri Lankan national weightlifting team and university sports teams. Each set contains three or five consecutive lifts by one participant. The protocol used 26 IMUs distributed across the limbs and spine, recording triaxial acceleration and angular velocity at 100~Hz. Frontal and sagittal video recorded at 30~frames per second provided a visual record for annotation. The video and inertial streams were aligned offline, and synchronised videos are included for selected sets. Three annotators independently identified 12 phase markers for each lift. The intervals between successive markers define 11 phases, including preparation, pulling, recovery, overhead stabilisation and return to rest. Both consensus boundaries and individual annotations are available.

The combination of repeated lifts, known loads and anatomical sensor locations supports comparisons within a participant as well as evaluation on previously unseen participants. For example, a phase-segmentation method can be tested on whether it detects the transition from front rack rest to the jerk, and its timing error can be compared with the spread of the three human annotations at the same boundary. The sensor layout also allows researchers to select a smaller set of anatomical sites and quantify how much phase information those sites retain, a question relevant to sparse inertial motion capture \cite{yi2025}. Documented gaps and missing channels provide a separate opportunity to test performance under the sensor losses encountered during actual lifting sessions.

\newpage

\section{Methods}
\label{sec:methods}

\subsection{Study design and participants}

The study recorded athletes performing the Olympic weightlifting clean and jerk using body-mounted inertial measurement units (IMUs). Data collection took place at the University of Peradeniya Gymnasium. Athletes were recruited through the university weightlifting team, national weightlifting team of Sri Lanka, and other sports teams of the University of Peradeniya. Training experience was documented using a survey which all participants had to fill.

The working definitions of this dataset and the methodology are as explained in table \ref{tab:working_definitions}. Moreover, the definitions of the 12 phase markers are explained sequentially in figure \ref{fig:phase_markers}

\begin{table}[htbp]
    \centering
    \small
    \caption{Working definitions used in the dataset.}
    \label{tab:working_definitions}
    \renewcommand{\arraystretch}{1.2}

    \begin{tabularx}{\linewidth}{
        @{}l >{\raggedright\arraybackslash}X@{}
    }
        \toprule
        \textbf{Term} & \textbf{Definition} \\
        \midrule

        Participant &
        An athlete who performs clean and jerk lifts for
        data collection. \\

        Metadata &
        Participant anthropometrics and background,
        including age, sex, height, body mass, training
        experience and performance background. \\
        
        Session &
        A particular day of data collection. A session may
        contain several sets. \\

        Set &
        A collection of three or five clean and jerk lifts
        performed consecutively by the same participant. \\

        Lift &
        A single clean and jerk lift. \\

        Phases &
        The 11 temporal movement segments of a lift. \\

        Phase markers &
        The 12 annotated time points defining the boundaries
        of the 11 phases, each represented by a timestamp
        and a video frame index. Each marker denotes the
        start of its phase, except \texttt{end\_rep},
        which marks the end of the final phase
        (\texttt{end\_rest}) and the lift. \\

        \bottomrule
    \end{tabularx}
\end{table}



\subsection{Ethics statement and consent}


The Ethical Review Committee of the Faculty of Engineering, University of
Peradeniya, Sri Lanka, approved the study (approval no.~ERC/FoE/2025/001).
All participants provided written informed consent to participate.


The direct participant identifiers have been removed for added privacy through de-identification in the released version. In the videos, the faces of the participants have been blurred out. Moreover, in the participant metadata, the names of the participants are removed. 

\subsection{Instrumentation and sensor placement}

Each IMU recorded acceleration and angular velocity along three local sensor axes. The inertial recordings were acquired at 100~Hz using the in-house developed wearable device and firmware by Samarasekera et~al.\cite{samarasekera2026}. The accelerometer and gyroscope were configured to \(4 \times 9.81\,\mathrm{m\,s^{-2}}\) and 1000~deg/s ranges respectively. 

\begin{figure}[htbp]
\centering
\includegraphics[width=\linewidth]{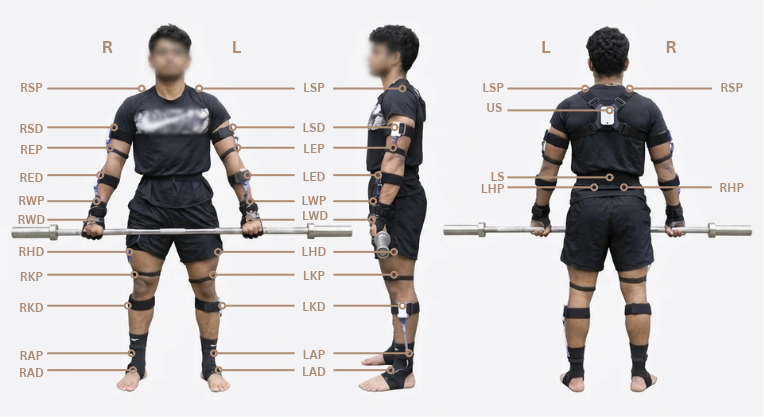}
\caption{Tri-view (anterior, left lateral, and posterior) illustration of the 26-IMU sensor placement. The 24 bilateral limb IMUs follow the standardised \texttt{\{L/R\}\{S/E/W/H/K/A\}\{P/D\}} convention for left/right (\texttt{L}/\texttt{R}), joint (shoulder \texttt{S}, elbow \texttt{E}, wrist \texttt{W}, hip \texttt{H}, knee \texttt{K}, ankle \texttt{A}), and proximal/distal (\texttt{P}/\texttt{D}) positions, complemented by two spinal IMUs at the upper spine (\texttt{US}) and lower spine (\texttt{LS}).}
\label{fig:sensor_placement}
\end{figure}

Table~\ref{tab:sensor_placement} and figure \ref{fig:sensor_placement} explain the sensor placement of IMUs  throughout the body of participants. Additionally, figure \ref{fig:sensor_placement} also shows axis directions of sensors in each of the mounting locations. 

\begin{table}[htbp]
\centering
\renewcommand{\arraystretch}{1.15}
\setlength{\tabcolsep}{10pt}
\caption{Standardised nomenclature of the 26 body-mounted IMUs grouped by anatomical region.}
\label{tab:sensor_placement}
\begin{tabular}{@{}lcccc@{}}
\toprule
& \multicolumn{2}{c}{\textbf{Left}} & \multicolumn{2}{c}{\textbf{Right}} \\
\cmidrule(lr){2-3} \cmidrule(lr){4-5}
\textbf{Joint / Region} & \textbf{Proximal} & \textbf{Distal} & \textbf{Proximal} & \textbf{Distal} \\
\midrule
Shoulder & \texttt{LSP} & \texttt{LSD} & \texttt{RSP} & \texttt{RSD} \\
Elbow    & \texttt{LEP} & \texttt{LED} & \texttt{REP} & \texttt{RED} \\
Wrist    & \texttt{LWP} & \texttt{LWD} & \texttt{RWP} & \texttt{RWD} \\
\addlinespace[3pt]
Hip      & \texttt{LHP} & \texttt{LHD} & \texttt{RHP} & \texttt{RHD} \\
Knee     & \texttt{LKP} & \texttt{LKD} & \texttt{RKP} & \texttt{RKD} \\
Ankle    & \texttt{LAP} & \texttt{LAD} & \texttt{RAP} & \texttt{RAD} \\
\midrule
Spine (midline) & \multicolumn{2}{c}{\texttt{US} (upper)} & \multicolumn{2}{c}{\texttt{LS} (lower)} \\
\bottomrule
\end{tabular}
\end{table}

\begin{figure}[htbp]
\centering
\includegraphics[width=0.85\linewidth]{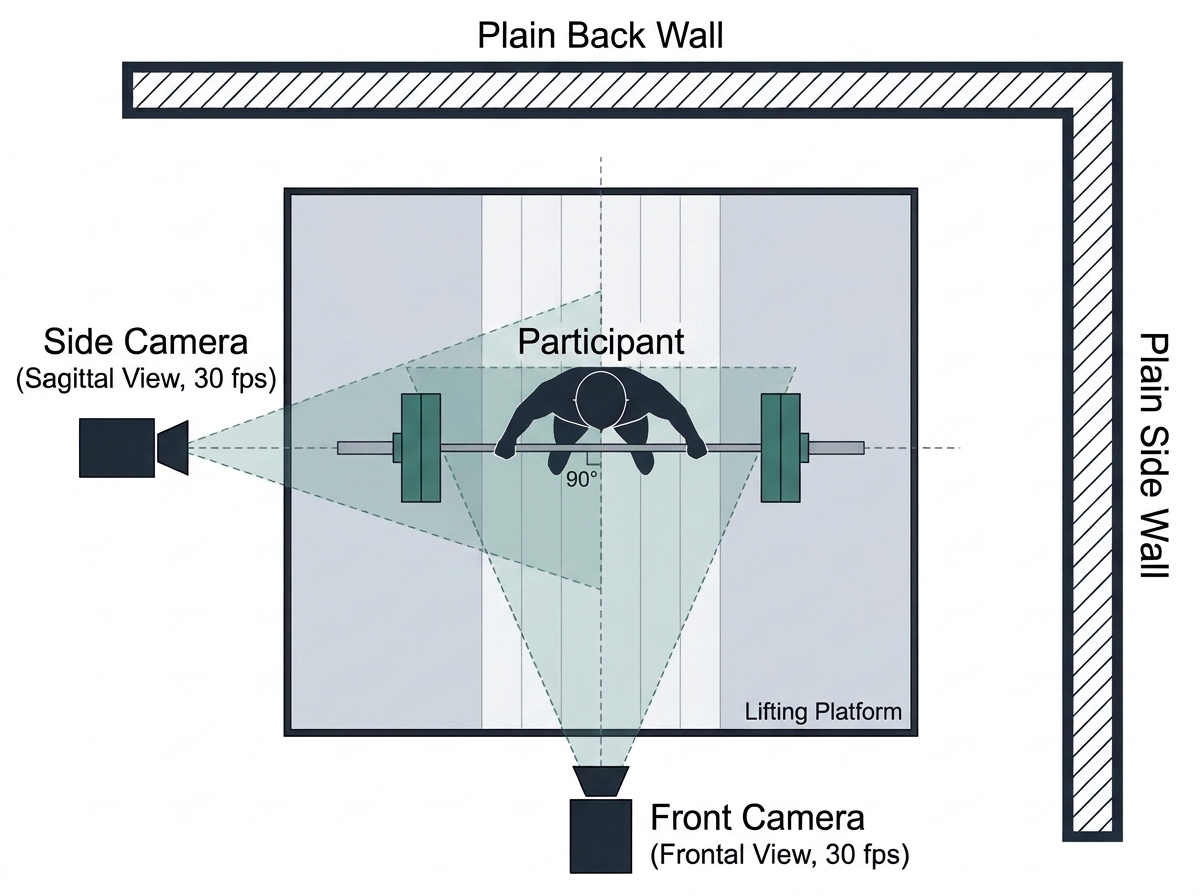}
\caption{Top-down schematic of the dual-camera weightlifting data-acquisition layout, showing the orthogonal frontal and sagittal camera views (30~fps) centred on the lifting platform against plain background walls.}
\label{fig:camera_placement}
\end{figure}

The video footage was taken at 30~fps and 1080p resolution by placing cameras on 2 locations. The locations were selected such that the imaginary straight lines drawn from the participant to the cameras were perpendicular to each other (i.e. one camera capturing the front view and one camera capturing the side view) as shown in figure \ref{fig:camera_placement}. The perpendicularity was made sure by using a long straight edge and vertical walls close to the lifter. 

\subsection{Experimental protocol}

\begin{figure}[h]
\centering
\includegraphics[width=\linewidth]{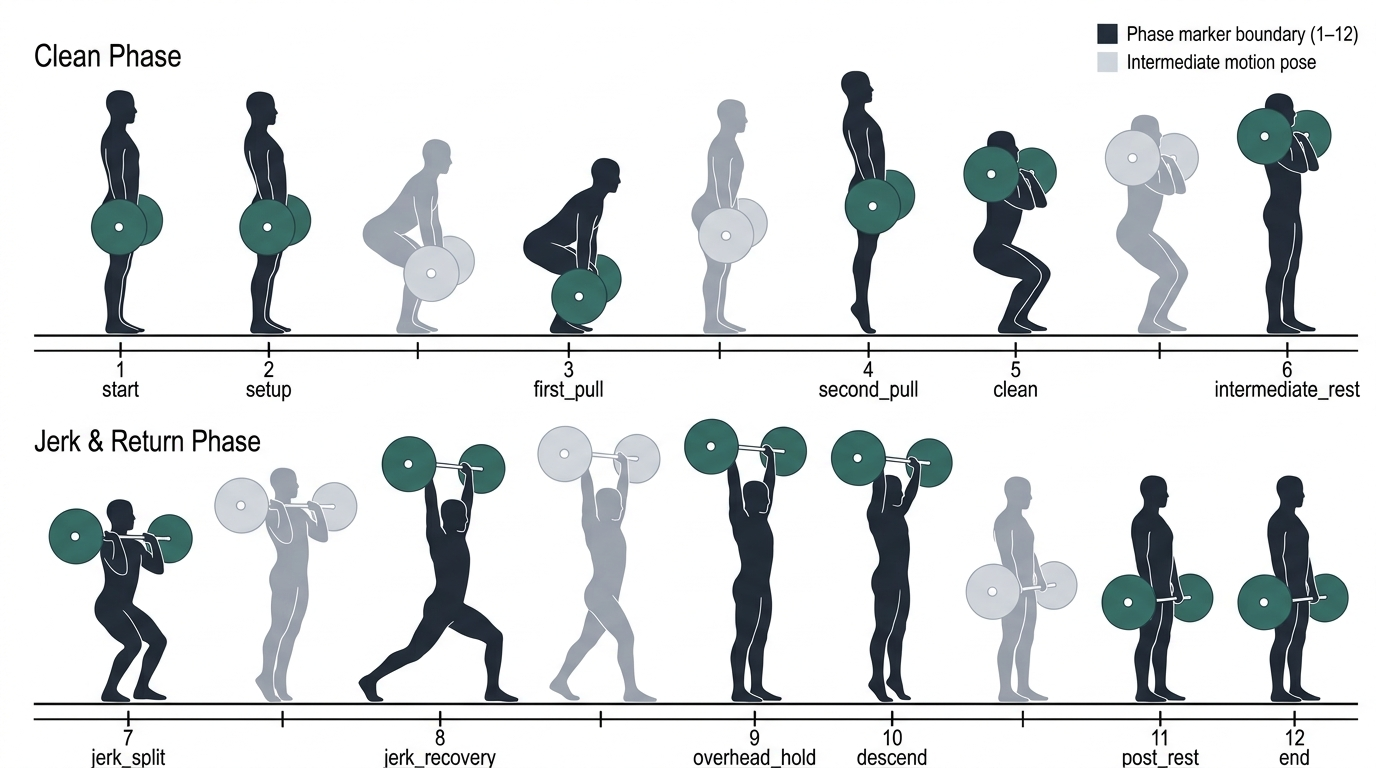}
\caption{Side-profile kinematic sequence of a Clean \& Jerk lift illustrating the 12 standardised phase-marker boundaries (dark charcoal, numbered 1--12) and intermediate in-phase motion poses (light grey) across the clean (top) and jerk-and-return (bottom) stages.}
\label{fig:phase_markers}
\end{figure}

In each clean and jerk lift, the athlete lifted the barbell from the floor to the shoulders in the clean and then from the shoulders to an overhead position in the jerk. Figure \ref{fig:phase_markers} shows the different stages (phase markers) of each lift. Before the lifting bouts, a participant completed warm-up, and then the device was worn. Afterwards the participant performed a few more warm up exercises and practice clean and jerk lifts to see if the device is well fit and if he/she is comfortable to use it. Finally, sensor checks were performed to see if all physical and wireless connections were in good condition before data collection began. The acquisition setup used either a 15kg barbell or a 20kg barbell with or without 5kg or 10kg weights on either side depending on the set. There were certain participants who wore footwear and certain participants who did not. This choice was based on each participant's preference and the ability to place the wearable device's sensor on the foot of the participant. All data collection sessions were conducted with a coach near by to supervise if any accidents or injuries. Loads were selected based on the athlete's ability to lift the load the coach's expert advise. The starting load was 20~kg, with a reduced load of 15~kg used when a participant was uncomfortable lifting 20~kg or when recommended by the coach. Participants who lifted 20~kg went on to lift 30~kg and 40~kg if the participant and the coach were both comfortable with it. Participants were instructed to stay still before the first lift(start\_rest), after the jerk phase of each lift and after bringing the bar down to the start position upon completion each lift(end\_rest). Rest and refreshments were provided between each set of the same participant to maintain their conditioning. 


\subsection{Data Screening and Synchronisation}

\subsubsection{Data Screening Process}
A total of 60 sets were collected and screened for inclusion in the released dataset. Sets were excluded when an incident, such as an accidental barbell drop, led the coach to advise discontinuation, or when connection issues resulted in data loss or corruption across multiple sensors. Out of a total of 60~sets, six sets were excluded on these grounds, leaving 54 sets in the released dataset. Furthermore, the released dataset only contains front and side videos for 28~sets. This reduced number is either due to those participants not agreeing to release their video publicly or due to privacy issues arising from the presence of bystanders. 

\subsubsection{Multi Model Data Synchronisation}
During data acquisition, the frontal and sagittal videos were recorded independently by independent operators. Before using videos as reference to the inertial measurements, a time synchronisation of multi model data was done offline using a movement energy cross-correlation pipeline as explained here.

The information on both videos was extracted in the form of two-dimensional pixel position of 12 body key points $\mathcal{V}_n$ (shoulders, elbows, wrists, hips, knees and ankles) per each frame $n$ using YOLOv11x-pose (yolo11x-pose.pt) \cite{yolo11_ultralytics} and  low confidence estimations ($c_{j,n}  < 0.3  j \in \mathcal{V}_n$) were masked and  linearly interpolated across frames. The one-dimensional movement energy signal for the $n^{th}$ frame of each video $E_{vid}[n]$, was constructed by summing the frame-to-frame pixel displacements across the key points $\mathcal{V}_n$. Where $\Delta x_{j,n} = x_{j,n} - x_{j,n-1}$ and $\Delta y_{j,n} = y_{j,n} - y_{j,n-1}$ denote the horizontal and vertical pixel displacements of key point $j$ between consecutive frames. The horizontal pixel displacements was down weighted by the factor of $0.25$ to emphasise the shared vertical motion visibility across both camera views.

\begin{equation}
E_{\text{vid}}[n] = \sum_{j \in \mathcal{V}_n} \sqrt{0.25\,\left(\Delta x_{j,n}\right)^2 + \left(\Delta y_{j,n}\right)^2},
\label{eq:video_energy}
\end{equation}

Similarly, The one-dimensional movement energy signal for $m^th$ sample of the inertial measurements $E_{\text{IMU}}[m]$, was constructed by summing the gravity compensated acceleration magnitude across all IMU sensors $\mathcal{S}$. Where, $\mathbf{a}_s[m] = (a_{x,s}[m],\, a_{y,s}[m],\, a_{z,s}[m])$ is the acceleration vector of sensor $s$ at sample $m$, $\|\mathbf{a}_s[m]\|_2$ is its Euclidean magnitude, and $g_s = \operatorname{median}(\|\mathbf{a}_s\|_2)$ is the static gravity baseline of sensor $s$ estimated across the full set. To match the video frame rate, $E_{\text{IMU}}$ was then resampled from 100Hz onto 30Hz.

\begin{equation}
E_{\text{IMU}}[m] = \sum_{s \in \mathcal{S}} \left|\,\|\mathbf{a}_s[m]\|_2 - g_s\right|,
\label{eq:imu_energy}
\end{equation}

\begin{figure}[htbp]
\centering
\includegraphics[width=\linewidth]{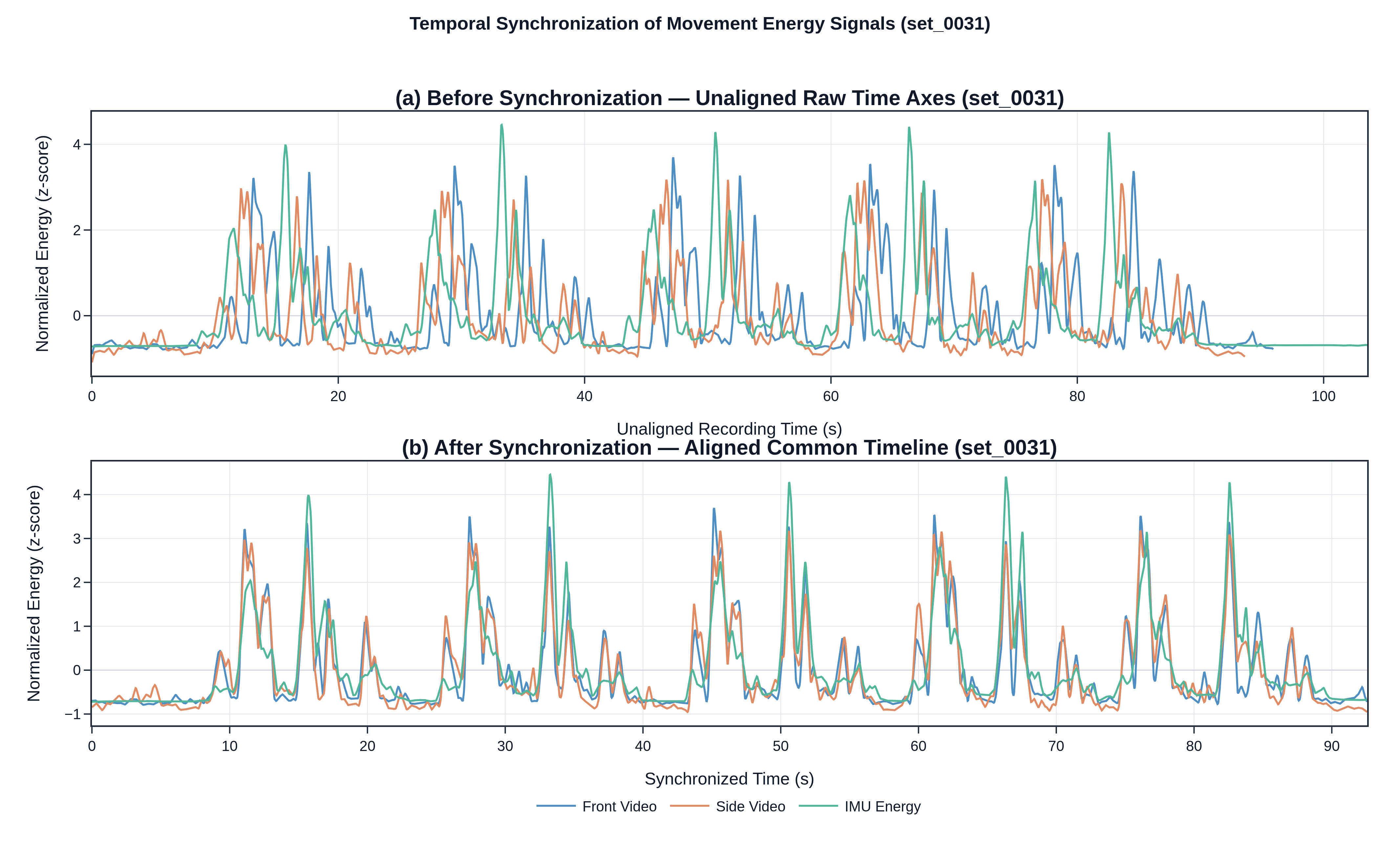}
\caption{Time synchronisation of the frontal video, sagittal video, and IMU movement-energy signals for \texttt{set\_0031}. (\textbf{a}) Before synchronisation. (\textbf{b}) After synchronisation onto a common timeline.}
\label{fig:sync_example}
\end{figure}

Finally, pairwise cross correlation was evaluated between the constructed movement-energy signals to determine the optimal time lags that maximised correlation between the frontal video, sagittal video, and inertial measurement streams. Multi model data were then aligned to a common timeline and trimmed to their shared time span as shown on (Fig.~\ref{fig:sync_example}).

\newpage
\section{Data Records}
\label{sec:data_records}

\begin{figure}[htbp]
\centering
\includegraphics[width=\linewidth]{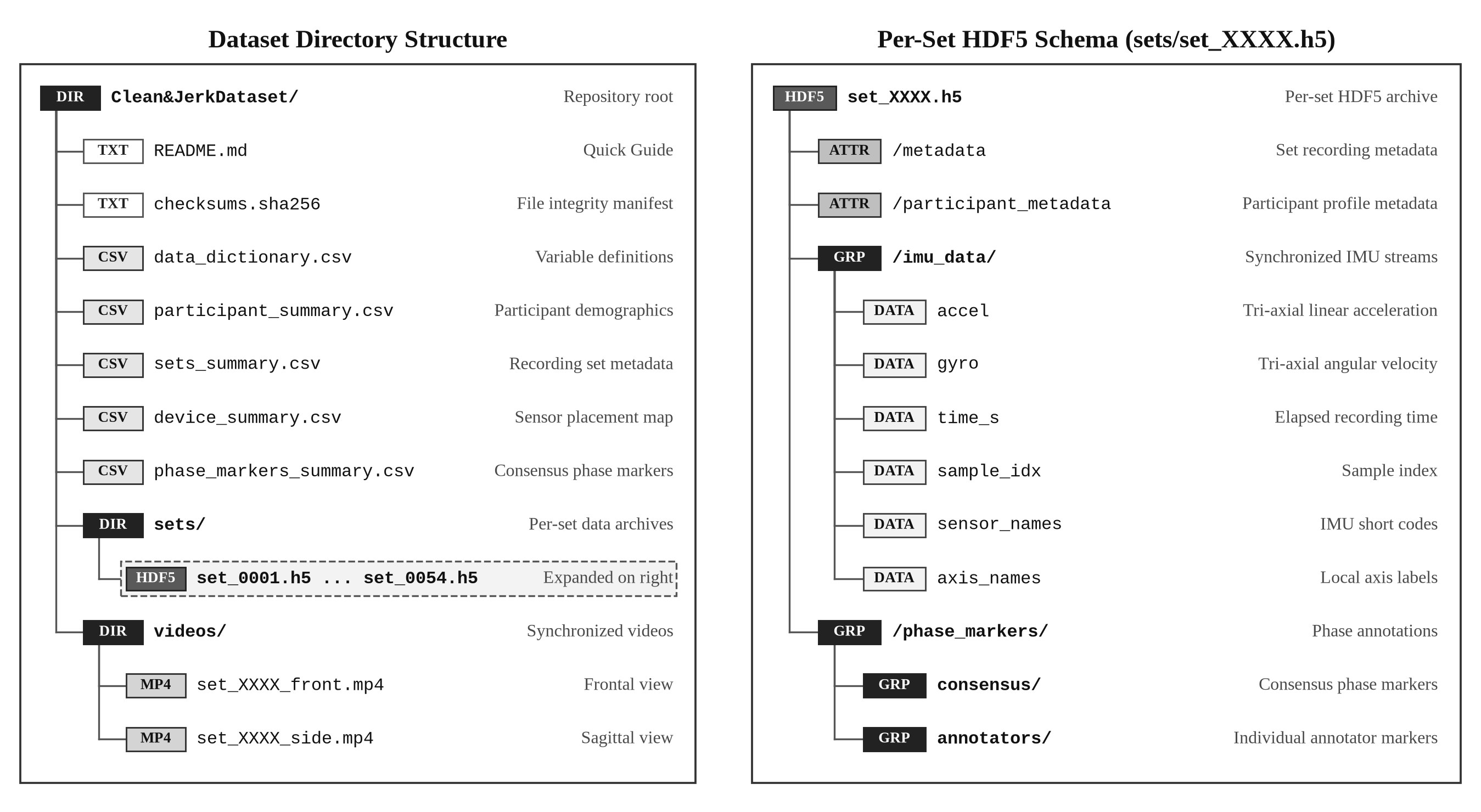}
\caption{Dataset directory structure and per-set HDF5 schema.}
\label{fig:folders}
\end{figure}

The complete dataset (available on Zenodo at \url{https://doi.org/10.5281/zenodo.24676946}) presented in this study is openly available on Zenodo under the CC BY 4.0 license. The dataset contains 54 recording sets, totalling 123~files and 8.1~GB. The directory structure is illustrated in Fig.~\ref{fig:folders}. The root directory contains five CSV files providing an overall summary of the dataset, accompanied by a \texttt{README.md} quick-start guide and \texttt{checksums.sha256} file integrity manifest. The complete data package for each set is archived in an HDF5 file and organized under the \texttt{/sets} directory. Additionally, a selected number of synchronized reference videos are organized under the \texttt{videos/} directory. 

\textbf{Summary files.} Five UTF-8 CSV files in the root directory provide a complete summary and index the data in the repository. The dataset is fully linked across tables using three primary keys: \texttt{set\_id}, \texttt{participant\_id}, and sensor \texttt{short\_code}. An overview of each summary file and its contents is provided in Table~\ref{tab:summary_files}.

\begin{table}[htbp]
\centering\scriptsize
\caption{Summary of root-level CSV files in the dataset repository.}
\label{tab:summary_files}
\begin{tabularx}{\linewidth}{@{}L{44mm}Y@{}}
\toprule
File & Contents \\
\midrule
\texttt{data\_dictionary.csv} & Field definitions, data types, and physical units for whole dataset. \\
\texttt{device\_summary.csv} & Mapping of each IMU short code to its anatomical location, body side, joint, and measurement specifications. \\
\texttt{participant\_summary.csv} & De-identified athlete demographics, body metrics, training background. \\
\texttt{sets\_summary.csv} & Per-set recording metadata, including participant link, timestamp, barbell load, lift counts, duration. \\
\texttt{phase\_markers\_summary.csv} & Phase markers per lift aligned across video frames, IMU sample indices, and timestamps, with inter-annotator agreement statistics. \\
\bottomrule
\end{tabularx}
\end{table}

\textbf{Set files.} Each recording set is packaged as a independent HDF5 archive (\texttt{<set\_id>.h5}) that can be inspected with standard HDF5 viewers or loaded directly using the Python utilities provided in the code repository. The internal group hierarchy is shown in Fig.~\ref{fig:folders}, and the corresponding array shapes, physical units, and data types are summarised in Table~\ref{tab:variables}. Set metadata, participant metadata and phase markers were grouped in the archive file allowing users to analyse each set independently. All inertial measurement channels are expressed in each individual sensor's local right-handed (X, Y, Z) coordinate frame. Any missing or unrecorded samples are represented as NaN, differentiating them from zero-valued data. Additionally, the \texttt{/phase\_markers/annotators} group preserves individual per-annotator records, allowing users to establish their own representations of inter-annotator variability and marker uncertainty.

\begin{table}[htbp]
\centering\scriptsize
\caption{Key fields of the per-set HDF5 files (\texttt{sets/set\_XXXX.h5}).}
\label{tab:variables}
\begin{tabularx}{\linewidth}{@{}Y L{24mm} L{34mm}@{}}
\toprule
Field & Shape & Unit / type \\
\midrule
\texttt{/metadata} (attrs) & -- & mixed \\
\texttt{/participant\_metadata} (attrs) & -- & mixed \\
\texttt{/imu\_data/accel} & $(N, 26, 3)$ & m/s$^2$ / \texttt{float32} \\
\texttt{/imu\_data/gyro} & $(N, 26, 3)$ & deg/s / \texttt{float32} \\
\texttt{/imu\_data/time\_s} & $(N,)$ & s / \texttt{float64} \\
\texttt{/imu\_data/sample\_idx} & $(N,)$ & samples / \texttt{int32} \\
\texttt{/imu\_data/sensor\_names} & $(26,)$ & \texttt{S4} (str) \\
\texttt{/imu\_data/axis\_names} & $(3,)$ & \texttt{S2} (str) \\
\texttt{/phase\_markers/consensus/*} & $(C,)$ each & frames, samples, s \\
\texttt{/phase\_markers/annotators/*} & $(M,)$ each & frames, samples, s \\
\bottomrule
\end{tabularx}
\end{table}

\textbf{Video files.} Time synchronised reference videos are stored in the \texttt{/videos/} directory, following the naming convention \texttt{<set\_id>\_<front/side>.mp4}. All videos are recorded in portrait orientation and encoded in H.264 MP4 format at a resolution of \(1080 \times 1920\) pixels and a frame rate of 30 fps for maximum compatibility. To protect participant privacy, faces are blurred in all released video files prior to publication.

\newpage
\section{Technical Validation}
\label{sec:technical_validation}

\subsection{Sensor accuracy and resolution}

\textbf{Prior Hardware Validation.} The Synchronized Multi-IMU Wearable System used in the study has been already validated on previous instrumentation studies Adhikari et al. \cite{adhikari2026} , Samarasekera et al.\cite{samarasekera2026}, and establish a basis for the reliability and accuracy. Adhikari et al. \cite{adhikari2026}  demonstrate the multi IMU architecture, precision timing, a real-time data logging capabilities of the system and validate against a video-based pose estimation. Extending on that Samarasekera et al.\cite{samarasekera2026} demonstrate on a 2 h 12 min static rigid-body elbow hold, system's drift rate is near zero \(1.5 \times 10^{-6}\) and angular resolution is \(0.6442^\circ\). 

\textbf{Data Completeness.} Initially a total of 60 sets were collected, 6 sets (10\%) was excluded from the dataset during screening process due to severe data incompleteness, resulting 54 released sets. Across all $54 \times 26 \times 6 = 8{,}424$ individual sensor-axis channels (26 six-axis IMUs per set), $8{,}094$ ($96.1\%$) was complete, $168$ ($2.0\%$) had partial gaps, and $162$ ($1.9\%$) were completely missing. 

\subsection{Multi Model Data Synchronisation}
Time synchronisation accuracy across all 54 sets was validated using two metrics on the normalised movement-energy signals. (i) Pearson correlation ($r$) before versus after alignment and (ii) local residual timing lag ($\Delta t_{\text{res}}$) across 10s sliding windows to verify timing accuracy in milliseconds.

\begin{figure}[htbp]
\centering
\includegraphics[width=\linewidth]{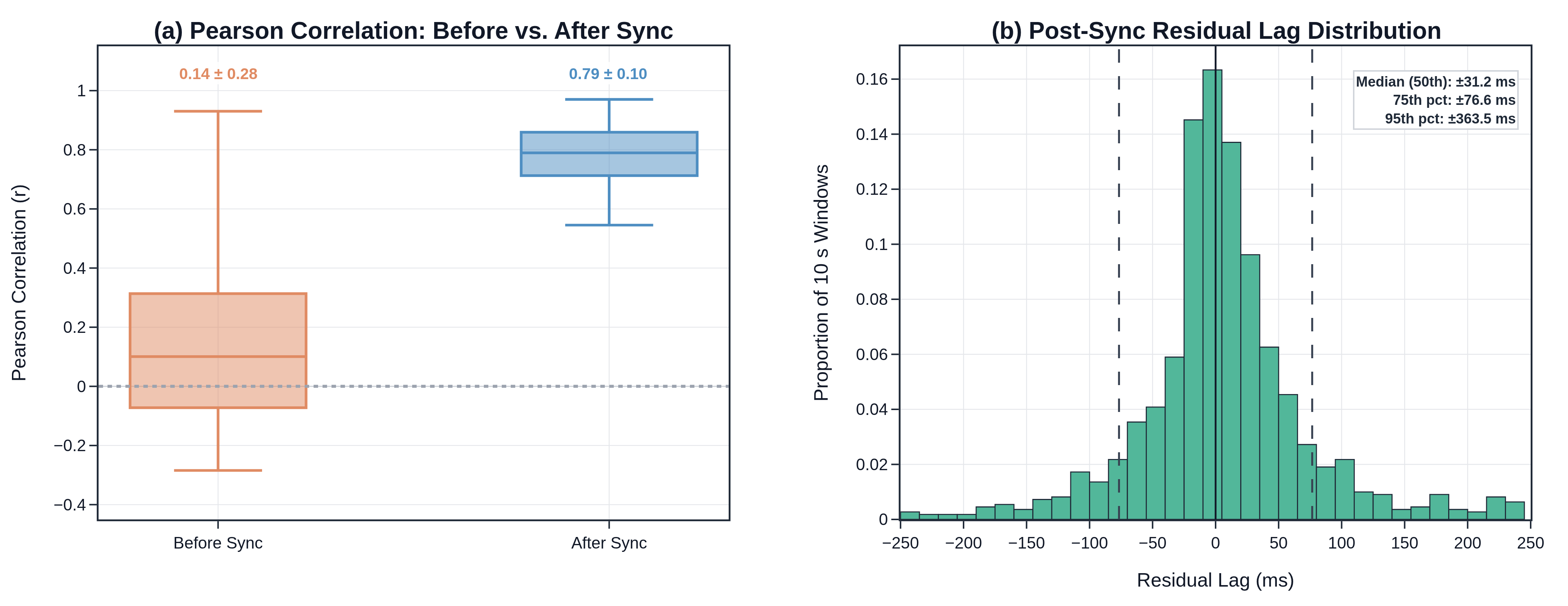}
\caption{Multimodal synchronisation accuracy across sets. (\textbf{a}) Pearson correlation before and after synchronisation. (\textbf{b}) Post-synchronisation residual lag distribution across 10~s windows.}
\label{fig:sync_results}
\end{figure}

In the whole dataset, these two metrics were evaluate between each video, IMU pairs and mean person correlation increased from $0.14 \pm 0.28$ to $0.79 \pm 0.10$ before and after synchronisation (Fig.~\ref{fig:sync_results}a). The residual timing lag was calculated after synchronisation for 10s sliding windows with a 5~s step (Fig.~\ref{fig:sync_results}b). Distribution centred on $0\ \text{ms}$ with a median absolute residual lag of $31.2\ \text{ms}$, a 75th percentile of $76.6\ \text{ms}$, and a 95th percentile of $363.5\ \text{ms}$. Resulting a  alignment precision which is well within acceptable region for phase-level Clean \& Jerk movement analysis.

\subsection{Annotation Agreement}
\label{subsec:annotation_agreement}

Inter-annotator reliability across the three independent raters for each set was evaluated over all phase markers by considering the sample standard deviation (SD) across the three annotated timestamps per each phase marker in each lift in all the sets.

\begin{figure}[htbp]
\centering
\includegraphics[width=\linewidth]{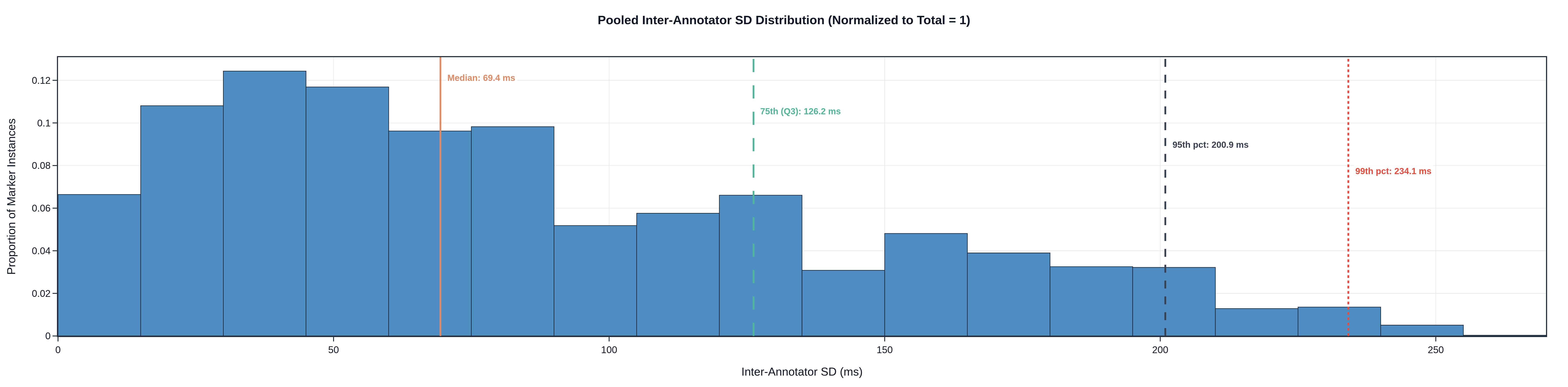}
\caption{Distribution of pooled inter-annotator standard deviations across all 2,952 phase-marker instances.}
\label{fig:sd_dist}
\end{figure}

All 2,952 marker instances ($100.0\%$) were identified by all three raters using reference videos. Across all markers (Fig.~\ref{fig:sd_dist}), the median of inter-annotator standard deviation was $69.4\ \text{ms}$, with $95\%$ of boundaries agreeing within $200.9\ \text{ms}$ ($\approx 6$ video frames) and all boundaries within $266.7\ \text{ms}$ (8 video frames).

As shown in Table~\ref{tab:agreement}, agreement was highest on sharp dynamic transitions (e.g., \texttt{first\_pull}, \texttt{second\_pull}, , and \texttt{clean}) and lowest on gradual rest transitions (e.g., \texttt{post\_rest} and \texttt{overhead\_hold}), where exact onset of stillness is naturally less distinct.

\begin{table}[htbp]
\centering\small
\caption{Inter-annotator agreement across all 12 standardised phase markers ($n = 246$ lifts per marker; values shown as ms and equivalent 30~fps video frames).}
\label{tab:agreement}
\begin{tabular}{@{}lccc@{}}
\toprule
Phase Marker & Median SD, ms (frames) & 95th pct SD, ms (frames) & Max SD, ms (frames) \\
\midrule
\texttt{start} & 69.4 (2.1) & 216.2 (6.5) & 250.2 (7.5) \\
\texttt{setup} & 83.9 (2.5) & 194.6 (5.8) & 240.4 (7.2) \\
\texttt{first\_pull} & 50.9 (1.5) & 170.2 (5.1) & 250.2 (7.5) \\
\texttt{second\_pull} & 50.9 (1.5) & 101.4 (3.0) & 183.6 (5.5) \\
\texttt{clean} & 69.4 (2.1) & 153.7 (4.6) & 189.5 (5.7) \\
\texttt{intermediate\_rest} & 98.1 (2.9) & 214.3 (6.4) & 251.7 (7.6) \\
\texttt{jerk\_split} & 66.7 (2.0) & 192.4 (5.8) & 236.5 (7.1) \\
\texttt{jerk\_recovery} & 83.9 (2.5) & 207.0 (6.2) & 250.2 (7.5) \\
\texttt{overhead\_hold} & 107.2 (3.2) & 232.3 (7.0) & 266.7 (8.0) \\
\texttt{descend} & 69.4 (2.1) & 183.6 (5.5) & 214.3 (6.4) \\
\texttt{post\_rest} & 117.1 (3.5) & 216.9 (6.5) & 250.2 (7.5) \\
\texttt{end} & 66.7 (2.0) & 171.1 (5.1) & 234.1 (7.0) \\
\midrule
Overall & 69.4 (2.1) & 200.9 (6.0) & 266.7 (8.0) \\
\bottomrule
\end{tabular}
\end{table}


\section{Usage Notes}
\label{sec:usage_notes}

Use \texttt{sets\_summary.csv} to select recording sets by participant or  barbell load, then open the corresponding HDF5 files using \texttt{set\_id}. Use the average phase markers to segment the IMU signals and the individual annotations to assess boundary uncertainty. Match available frontal and sagittal videos to the IMU recordings by \texttt{set\_id} and use the aligned timestamps to locate corresponding events. 

Treat \texttt{NaN} values as missing measurements and check channel completeness before analysis. For machine-learning evaluation, keep all sets from each participant in the same split to prevent data leakage. For analyses that depend on precise video-IMU timing, consider the residual alignment error reported in Technical Validation.







\section{Data Availability}
\label{sec:data_availability}

Version~1 of the MM-CleanJerk dataset is available on Zenodo at \url{https://doi.org/10.5281/zenodo.24676946} 
The public release includes inertial measurements from 54 sets of clean and jerk lifts, phase annotations for 246 lifts, and de-identified participant and set metadata. These files are available under the Creative Commons Attribution 4.0 International licence (CC~BY~4.0), which permits reuse, adaptation, and redistribution with appropriate attribution and an indication of any changes.



\section{Code Availability}
\label{sec:code_availability}

Python code for accessing MM-CleanJerk is available in the \texttt{code/} directory of version~1 of the dataset on Zenodo (\url{https://doi.org/10.5281/zenodo.24676946})
The module \texttt{dataset\_loader.py} reads the CSV summary tables and per-set HDF5 files, retrieves signals from selected IMUs, and extracts individual lifts with a phase label for each sample. The interactive Jupyter notebook \texttt{example\_usage.ipynb} demonstrates these operations using a recorded set. The accompanying \texttt{requirements.txt} lists the Python dependencies.



\bibliographystyle{naturemag}
\bibliography{references}

\end{document}